\documentclass[a4paper,11pt]{article}

\usepackage{amsmath}
\usepackage{amsfonts}
\usepackage{slashed}
\usepackage{dsfont}
\usepackage{verbatim} 
\usepackage{fullpage}
\usepackage{hyperref}
\hypersetup{
    colorlinks=true,
	linkcolor=black,
	citecolor=black,
	urlcolor=black,
	filecolor=black,
    pdfborder={0 0 0},
}

\DeclareMathAlphabet{\mathbf}{OT1}{pnc}{b}{n}

\newfont{\bbbold}{msbm10 scaled \magstep1}

\def\bb1{\mathds{1}}

\def\bbE{\mbox{\bbbold E}}

\def\bbR{\mbox{\bbbold R}}

\def\bbT{\mbox{\bbbold T}}

\def\bbZ{\mbox{\bbbold Z}}

\def\cA{{\cal A}}

\def\cD{{\cal D}}
\def\cE{{\cal E}}
\def\cF{{\cal F}}

\def\cH{{\cal H}}

\def\cL{{\cal L}}
\def\cM{{\cal M}}
\def\cN{{\cal N}}

\def\cR{{\cal R}}

\newfont{\goth}{eufm10 scaled \magstep1}

\def\a{\alpha}
\def\b{\beta}
\def\c{\gamma}		\def\C{\Gamma}
\def\ch{\chi} 					
\def\d{\delta}		\def\D{\Delta}		
\def\e{\epsilon}	
\def\f{\phi}			\def\F{\Phi}	        
\def\h{\eta}

\def\l{\lambda}		\def\L{\Lambda}		
\def\m{\mu}
\def\n{\nu}

\def\r{\rho}
\def\s{\sigma}		
\def\t{\tau}
		\def\Th{\Theta}			
\def\X{\Xi}
\def\x{\xi} 			
		\def\O{\Omega}

\def\z{\zeta}

\def\tld{\tilde}

\def\w{\wedge}

\def\xX{\times}

\def\la{\langle}
\def\ra{\rangle}
\newcommand{\ket}[1]{\left\vert #1 \right \ra}
\newcommand{\bra}[1]{\left\la #1 \right\vert}

\newcommand{\braket}[2]{\left \la #1 \vert #2 \right \ra}

\newcommand{\lra}{\longrightarrow}

\newcommand{\rep}[1]{{{#1}}}

\newcommand{\comm}[2]{{\big[ #1 , #2 \big]}}
\newcommand{\acom}[2]{{\big\{ #1 , #2 \big\}}}

\def\del{\partial}

\def\det{\text{det}\,}

\def\half{\frac{1}{2}}

\newcommand{\dop}{{\slashed\nabla}}

\newcommand{\pref}[1]{(\ref{#1})}

\renewcommand{\Re}{\text{Re}\,}
\renewcommand{\Im}{\text{Im}\,}

\numberwithin{equation}{section}
\title{}
\author{David Geissb\"uhler}

\begin{document} 
	\thispagestyle{empty} 
	
	\begin{flushright}

	\end{flushright}
	
	\vspace{35pt} 
	\begin{center}
		{ \Large{\bf Double Field Theory and $\cN=4$ Gauged Supergravity}}
		
		\vspace{35pt}
		
		\bf{David Geissb\"uhler}
		
		\vspace{15pt}
		
		\footnotesize{\tt geissbuehler@itp.unibe.ch}
		
		\vspace{35pt}
		
		{\it   Albert Einstein Center for Fundamental Physics\\
		Institute for Theoretical Physics, University of Bern\\
		Sidlerstrasse 5, CH-3012 Bern, Switzerland
		}
				
		\vspace{50pt}
		
		{ABSTRACT}
	\end{center}
	
\noindent
Double Field Theory describes the NS-NS sector of string theory and lives on a doubled spacetime. The theory has a local gauge symmetry generated by a generalization of the Lie derivative for doubled coordinates. For the action to be invariant under this symmetry, a differential constraint is imposed on the fields and gauge parameters, reducing their possible dependence in the doubled coordinates. We perform a Scherk-Schwarz reduction of Double Field Theory, yielding electric gaugings of half-maximal supergravity in four dimensions when integrability conditions are assumed. The residual symmetries of the compactified theory are mapped with the symmetries of the effective theory and the differential constraints of Double Field Theory are compared with the algebraic conditions on the embedding tensor. It is found that only a weaker form of the differential constraint has to be imposed on background fields to ensure the local gauge symmetry of the reduced action.
	
	\vspace{10pt} 
	
\baselineskip 6 mm

\section{Introduction}

Double Field Theory (DFT), introduced by Hull and Zwiebach in \cite{hzw09} (see also \cite{zwi11} for a review), is a field theory describing the massless modes of closed string field theory on a $2D$-dimensional doubled torus with coordinates $X^M = (x^i, \tld x_i)$, with $i= 1,...,D$. The field content consists of a tensor $\cE_{ij}(x,\tld x) = g_{ij} + B_{ij}$ and a generalized dilaton $d(x,\tld x)$, both depending on the usual coordinates and the dual ones. An action for $\cE_{ij}$ treated as a fluctuation around a flat toroidal background was constructed up to cubic order as a truncation of the full closed string field theory action. This theory has the interesting property that it is invariant under T-duality, $O(D,D)$ in the decompactified limit, acting linearly on the doubled coordinates $X^M$ and by M\"obius transformations on $\cE_{ij}$. Moreover, this theory has a local gauge symmetry \cite{hzw09b} generated by a generalized Lie derivative which is an extension for doubled coordinates of the Dorfmann bracket appearing in Generalized Complex Geometry \cite{gua03}. In DFT the dual coordinates have a physical relevance, even though the fields are constrained by the level-matching condition on the string state, which translates in a differential constraint for the massless modes
\begin{equation}\label{intro_wc}
	\del_i \tld \del^i \cE_{jk} = \del_i \tld \del^i d = 0.
\end{equation}
In \cite{hhz10} a background independent action for $\cE_{ij}$ was proposed. It was however found that, in this new formulation, a stronger differential constraint has to be imposed by hand in order to get the symmetries of the cubic theory. In a subsequent paper \cite{hhz10b}, an equivalent formulation of the theory in term of the generalized metric $\cH_{MN}$ was presented, clarifying the link between DFT and Generalized Complex Geometry in the context of type II flux compactifications \cite{gmpt04,gmrp05}. In the generalized metric version of the theory the field $\cE_{ij}$ is replaced by a symmetric $O(D,D)/O(D)\xX O(D)$ coset element, the generalized metric $\cH_{MN}$, satisfying 
\begin{equation}\label{intro_oddc}
    \cH_{MN}\h^{NP}\cH_{PQ}\h^{QR} = \cH_{MN}\cH^{NR}=\d_M^R.
\end{equation}
where $\h^{MN}$ is the off-diagonal $O(D,D)$ invariant metric. The advantage of this formulation is that T-duality acts linearly on $\cH_{MN}$ and leaves the generalized dilaton $d$ invariant. As in the background independent theory, all fields and gauge parameters have to satisfy this new \textit{strong constraint}, stronger than the one needed by level-matching. In addition to \pref{intro_wc}, the $O(D,D)$ invariant laplacian is also required to vanish on all products of fields and gauge parameters $\z^M$
\begin{equation}\label{intro_sc}
	\del_M\del^M \cH_{NP} = \del^M \cH_{NP}\, \del_M \cH_{QR} =\del^M \cH_{NP}\, \del_M d 
	=\del^M \cH_{NP}\, \del_M \z^Q = \del^M d\, \del_M d = ...= 0,
\end{equation}
where the derivative with an upper index is $\del^M= \h^{MN}\del_N$ such that $\del_M\del^M = 2\del_i \tld \del^i$. While the constraint \pref{intro_wc} states that each field propagates on the $O(D,D)$ lightcone, its stronger version enforces them to propagate in the same null isotropic subspace. This allows one, at least locally, to rotate the theory in a duality frame where the fields' dependence in the dual coordinates is gone. When there is no dependence in the $\tld x_i$ coordinate, the DFT action is equivalent to the standard action for the NS-NS sector of string theory. It however remains possible that there exists a generalization of this theory not requiring this strong constraint. When the theory is compactified, the original symmetries are broken up to a subgroup of residual symmetries which are symmetries of the effective theory. In this paper we will show for a Scherk-Schwarz reduction of DFT that a part of the strong constraint can be relaxed without affecting this residual symmetry.

Double Field Theory has recently been extended to include the gauge fields of the heterotic theory \cite{hkw11b} and the R-R fields of type II theories \cite{hkz11, hkz11b}. 
In \cite{hkw11}, by relaxing part of the strong constraint in the R-R sector, it was discovered that a modification of DFT yielding massive type IIA can be constructed, preserving the full gauge invariance of the theory. 
Notions of differential geometry compatible with the constraints on the generalized metric were developed in \cite{jlp10,jlp11} and couplings to fermions were worked out using this framework \cite{jlp11b}. In \cite{hkw10}, a frame-like reformulation of the theory with an explicit left/right factorization was constructed and connected to Siegel's early work on doubled theories \cite{sie93,sie93b}.

The doubled formalism already attracted much attention for its ability to account for non-geometric fluxes \cite{hul04,hre09,dpst07,dpr08,adp09}. While Scherk-Schwarz reductions \cite{scs79} with internal three-form flux \cite{kmy99} can only explain geometric fluxes that appear in the effective gauged supergravity, reductions with duality twists \cite{dhu02} and reductions on twisted doubled tori \cite{hre09} give also rise to non-geometric fluxes $Q_a{}^{bc}$ and $R^{abc}$. Non-geometric compactifications in the context of $\cN=1$ flux compactifications have also been studied \cite{gmpw08}, where the various fluxes have been identified with the charges appearing in the Courant bracket of internal doubled vielbeins. Recent results with worldsheet theories, showed correspondence between the vanishing of worldsheet anomalies and existence of $\cN=4$ vacua \cite{dpr08,adp09}.

On the effective theory side, these twists, either geometric or not, correspond to gaugings of maximal or half-maximal supergravities \cite{sam08}. These gauged theories are deformations of the original supergravities where a subgroup $G$ of the rigid duality group, $E_{7,7}$ or $SL(2,\bbR)\xX SO(6,n)$ in four dimensions for $\cN=8$ or $\cN=4$ respectively, has been promoted to a local gauge symmetry. The duality group acts on vectors via the fundamental representation the symplectic group, $Sp(56)$ for $\cN=8$, or $Sp(12 + 2n)$ for $\cN=4$ with $n$ vector multiplets, and the scalar fields live in the coset spaces $E_{7,7}/SU(8)$ or $SL(2,\bbR)\xX SO(6,n)/SO(2)\xX SO(6)\xX SO(n)$. The gauging is realized by embedding the adjoint of $G$, one generator $X_I$ for each gauge field and magnetic duals, in the duality group with generators $(T_\L)_I{}^J$ acting in the fundamental of the symplectic group, via the so-called embedding tensor \cite{wst05}
\begin{equation}\label{intro_et}
	X_I = \Th_I{}^\L T_\L.
\end{equation}
The embedding tensor is subject to two quadratic constraints, the first ensuring that the right number of gauge fields propagate and the second that the algebra closes
\begin{equation}\label{intro_gadj}
	\comm{X_I}{X_J} = -X_{IJ}{}^K X_K.
\end{equation}
A linear constraint is imposed such that the symplectic metric is invariant under $G$. A further linear constraint, related to gauge anomalies \cite{rstp08}, has to be imposed and is known from $\cN=8$ supersymmetry where it keeps only the $\rep{912}$ in the embedding tensor. Truncation of the $\cN=8$ theory to $\cN=4$ is done by embedding $SL(2,\bbR)\xX O(6,6)$ in $E_{7,7}$ and keeping only components that are even under a $\bbZ_2$ symmetry. The only representations that survive the constraints on the embedding tensor are a product of a fundamentals of $SL(2,\bbR)$ and $SO(6,6)$, $\x_{\a A}$ with $\a=(+,-)$, and a fundamental of $SL(2,\bbR)$ times an $SO(6,6)$ three index antisymmetric tensor, $f_{\a ABC}$ \cite{swe06,dgr11}. Electric gaugings corresponds to vanishing $f_{-ABC}$ and $\x_{-A}$ and in this special case the closure constraint reduces to
\begin{equation}\label{intro_etcs}
\begin{split}
    \x_+{}^A \x_{+A} &= 0\\
    \x_+{}^A f_{+ABC}&= 0\\
    f_{+[AB}{}^E f_{+C]DE}&= \frac{2}{3}f_{+[ABC}\x_{+D]},
\end{split}
\end{equation}
where indices have been raised and lowered with the invariant $SO(6,6)$ metric $\h_{AB}$.

In this paper we are interested to find a link between a class of DFT compactifications and electric gaugings of $\cN=4$ supergravity. To achieve this, we first reformulate the generalized metric version of DFT with a doubled vielbein. This formulation has a local $O(1,9)\xX O(1,9)$ Lorentz symmetry, which makes it equivalent to the construction of \cite{csw11}. The dynamics rely on the antisymmetric part $F_{ABC}$ and the trace part $\tld\O_A$ of an object $\O_{ABC}$ transforming as a spin connection under $O(1,9)\xX O(1,9)$. Upon Scherk-Schwarz reduction with a warp factor, the internal part of these objects get identified with $f_{+ABC}$ and $\x_{+A}$ and the theory reduces to $\cN=4$ gauged supergravity. We show that, together with integrability conditions, the strong constraint implies the closure relations \pref{intro_etcs} while the converse is not true in general. 

During the completion of this work another closely related work appeared \cite{abmn11}. Our results are slightly more general as we are able to find the general case of non-unimodular gaugings with an arbitrary internal dilaton. If we accept that the strong constraint can be relaxed in the compactified theory, for instance in a modified version of the theory, we find new solutions that do not enter the class of solutions found in \cite{abmn11}. 

Section 2 quickly reviews the ingredients of Double Field Theory and reformulate them with a doubled vielbein. A spin connection is defined and then used to rewrite the Dirac operator acting on R-R fields. In Section 3, half-maximal four-dimensional gauged supergravity is reviewed and dualized to get a propagating two-form. A Scherk-Schwarz reduction of DFT is then done and constraints analyzed. Finally, Section 4 gives a few examples of doubled backgrounds with and without strong constraint.

\section{Review of Double Field Theory and vielbien formulation}

A closed string compactified on a torus has quantized momenta $p_i$ and winding numbers $w^i$, with $i=1,...,D$. The possible values of the winding numbers and momenta are restricted by the level-matching condition 
\begin{equation}
    w^i p_i = \half \h^{MN} P_M P_N = N -\bar N
\end{equation}
where $N$ and $\bar N$ are the eigenvalues of the left/right number operators acting on the string state, $P_M = (p_i, w^i)$ and where $\h^{MN}$ is the off-diagonal $O(D,D)$ metric
\begin{equation}\label{dft_eta}
	\h = \left(\h^{MN}\right)
		= \begin{pmatrix}
			\h^{ij} & \h^i{}_j \\
            \h_i{}^j & \h_{ij}\\
		  \end{pmatrix}
		= \begin{pmatrix}
			0 & \d^i{}_j \\
			\d_i{}^j & 0 \\
		  \end{pmatrix}.
\end{equation}
If the torus has a constant metric $g_{ij}$ and a constant Kalb-Ramon two-form flux $B_{ij}$ on it, the Hamiltonian has a term depending of the windings and momenta of the form $P_M P_N \cH^{MN}$ where 
\begin{equation}\label{dft_gm}
	\cH = \left(\cH_{MN}\right)
		= \begin{pmatrix}
			\cH_{ij} & \cH_i{}^j\\
			\cH^i{}_j & \cH^{ij} \\
		  \end{pmatrix}
		= \begin{pmatrix}
			g_{ij} - B_{ik}g^{kl}B_{lj} & B_{ik}g^{kj} \\
			-g^{ik}B_{kj}		& g^{ij}\\
		  \end{pmatrix}.
\end{equation}
and $\cH^{MN} = \h^{MP}\h^{NQ}\cH_{PQ}$. T-duality along a coordinate $x^i$ acts by exchange of the associated winding and momentum $w^i \leftrightarrow p_i$. These transformations assemble with the discrete shift symmetry of the two-form to form the group $O(D,D,\bbZ)$, acting on momenta and windings as 
\begin{equation}
	P_M \lra O_M{}^N P_N, 
    \quad O_M{}^N \in O(D,D,\bbZ),
    \quad O_M{}^P \h_{PQ} O_N{}^Q = \h_{MN} 
\end{equation}
Under this transformation the level-matching condition is invariant because of the contraction with the $O(D,D)$ metric while $\cH_{MN}$ is redefined as $\cH_{MN}' = \cH_{PQ}O^P{}_M O^Q{}_N$ where $O^M{}_N = \h^{MP} O_P{}^Q \h_{QN}$. The dilaton is not invariant under T-duality, however, together with the determinant of the metric an invariant combinaison is
\begin{equation}\label{dft_dil}
	e^{-2d} = e^{-2\f}\sqrt{\vert g\vert}.
\end{equation}
While the coordinate $x^i$ is the conjugate of the momentum $p_i$ there is no obvious conjugate for the winding number. Double Field Theory \cite{hzw09}, inspired by closed string field theory, introduce new dual coordinates $\tld x_i$ that are conjugates of windings. The doubled coordinate $X^M = (x^i, \tld x_i)$ then transforms as an $O(D,D)$ vector under T-duality. The fields associated to the massless modes of the string are the metric $g_{ij}$, the Kalb-Ramon two-form $B_{ij}$ and the dilaton $\f$. They depend on both coordinates $x^i$ and $\tld x_i$, and enter the theory through the tensor $\cE_{ij}(x, \tld x) = g_{ij} + B_{ij}$ and through the invariant generalized dilaton $d(x, \tld x)$ defined in $\pref{dft_dil}$. The theory has local a symmetry generated by a pair of parameters $(\e_i, \tld\e_i)$ that also have a dependence in both coordinates. The level-matching condition then gives constraints on the the massless fields and parameters
\begin{equation}\label{dft_wc}
	\del_i\tld\del^i \cE_{jk} = \del_i\tld\del^i d  =
    \del_i\tld\del^i \e_j = \del_i\tld\del^i \tld\e_j =  0.
\end{equation}
In \cite{hzw09}, $\cE_{ij}$ was treated as a small fluctuation around a background and these level-matching conditions were sufficient for the theory to have the local gauge symmetry generated by $(\e_i, \tld\e_i)$. A background independent version was then found in \cite{hhz10}, with a stronger constraint on the fields.

The background independent version of this theory was reformulated in \cite{hhz10b} with the generalized metric \pref{dft_gm} as dynamical field. The action is
\begin{equation}\label{dft_action}
\begin{split}    
    S = \int d^{2D}X\,e^{-2d}\bigg( 
        \frac{1}{8}\cH^{MN} \del_M \cH^{PQ} \del_N  \cH_{PQ} 
		-\frac{1}{2}\cH^{MN} \del_M \cH^{PQ} \del_Q  \cH_{PN}&\\
		- 2  \cH^{MN}\del_M d\, \del_N d 
		+ 4 \del_M \cH^{MN}\del_N d& \bigg).
\end{split}
\end{equation}
where $\cH^{MN} = \h^{MP}\h^{NQ}\cH_{PQ}$ and where the fields depend on the doubled coordinate $X^M$. Since all indices are contracted with the $O(D,D)$ metric $\h_{MN}$ the action is manifestly invariant under $O(D,D)$, which is broken to the T-duality subgroup $O(D,D,\bbZ)$ when compactified on a torus. The expression \pref{dft_gm} is not invariant under $O(D,D)$, an invariant formulation can be achieved by forgetting this particular form and treating $\cH_{MN}$ as a symmetric $O(D,D)$ matrix
\begin{equation}
    \cH_{MN}\,\h^{NP}\,\cH_{PQ} = \h_{MQ}.
\end{equation}
The action \pref{dft_action} is also invariant under a local symmetry parameterized by an infinitesimal vector $\z^M(X)$ provided a constraint is imposed on the fields and parameters. This constraint originates the level-matching condition \pref{dft_wc} with the difference that the $O(D,D)$ laplacian $\del_M\del^M$ is also required to vanish when acting on every product of fields and parameters
\begin{equation}\label{dft_sc}
\begin{split}
    \del_M\del^M \cH_{NP} = \del_M\del^M d = \del_M\del^M \z^{N} =
    \del_M \cH_{NP}\,\del^M\cH_{QR} =  
    \del_M \cH_{NP}\,\del^M d =&\\
    \del_M \cH_{NP}\, \del^M\z^Q = \del_M d\,\del^M d =
    \del_M d\,\del^M \z^N = \del_M\z^N\,\del^M\z^Q =& 0.
\end{split}
\end{equation}
This \textit{strong constraint} tells that that the theory is not truly doubled, but written in an $O(D,D)$ covariant form. The variation of the generalized metric under the local symmetry is
\begin{equation}\label{dft_symm}
    \hat\d \cH_{MN} = \hat\cL_\z\cH_{MN}=\z^P\del_P \cH_{MN}
           + \left(\del_M \z^P - \del^P \z_M\right)\cH_{PN}
           + \left(\del_N \z^P - \del^P \z_N\right)\cH_{MP},
\end{equation}
where all indices are raised and lowered with $\h^{MN}$. This variation is generated by a generalized Lie derivative $\hat\cL$ that treats covariant and contravariant indices on the same footing. This modified Lie derivative vanishes when acting on $\h^{MN}$ such that it keeps intact the $O(D,D)$ structure of the theory. These transformations are called generalized diffeomorphisms and form an algebra when the strong constraint is assumed. The commutator of two such transformations with parameters $\z_{1,2}^M$ is again a generalized diffeomorphism with parameter $\z_3^M$ given by the C-bracket of $\z_{1,2}^M$
\begin{equation}
    \z_3^M = \comm{\z_1}{\z_2}^M_{(C)}
           = 2 \z_{[1}^N\del_N\z_{2]}^M
           - \z_{[1}^N\del^M \z_{2]N},
\end{equation}
which reduces to the Courant bracket for $\tld\del^m ...= 0$. For the action to be invariant, the dilaton pre-factor is required to transform as a density
\begin{equation}
	\hat \d e^{-2d} = \del_M\left(\z^Me^{-2d}\right).
\end{equation}
After some partial integrations the Lagrangian \pref{dft_action} can be rewritten as a quantity transforming as a scalar under \pref{dft_symm} 
\begin{equation}\label{dft_raction}
	S = \int d^{2D}X\,e^{-2d} \cR.
\end{equation}
This scalar $\cR$ also comes from the dilaton equation of motion $\cR=0$ and is called generalized Ricci scalar 
\begin{equation}
\begin{split}
	\cR = 
		\frac{1}{8}\cH^{MN} \del_M \cH^{PQ} \del_N  \cH_{PQ} 
		-\frac{1}{2}\cH^{MN} \del_M \cH^{PQ} \del_Q  \cH_{PN} 
		-\del_M\del_N \cH^{MN} &\\
		+4\cH^{MN}\del_M\del_N d 
		-4 \cH^{MN}\del_M d\, \del_N d 
		+ 4 \del_M \cH^{MN}\del_N d &
\end{split}
\end{equation}

\subsection{Vielbein formulation}

The generalized metric $\cH$ and the $O(D,D)$ metric $\h$ can be rewritten as squares of a doubled vielbein 
\begin{equation}\label{gf_gbeins}
\begin{split}
	\cH_{MN} &= {E^A}_M \cH_{AB} {E^B}_N, \\ 
	\h_{MN} &= {E^A}_M \h_{AB} {E^B}_N.
\end{split}
\end{equation}
Different gauge choices exist \cite{hhz10b}, for instance in \cite{sie93,sie93b,hkw10} a left/right $GL(D)\xX GL(D)$ symmetry is made manifest. We consider another option here, where $\h_{AB}$ is required to be numerically equal to $\h_{MN}$ and is used to raise and lower flat indices ($A,B,C...$). The second line in \pref{gf_gbeins} then constrains $E_A{}^M$ to be an element of $O(D,D)$.
$\cH_{AB}$ is chosen to have the $O(1,D-1)$ metric $h_{ab}=\text{diag}(-,+,..,+)$ in the diagonal blocks
\begin{equation}
    (\cH_{AB}) = \begin{pmatrix}
                    h_{ab} & 0\\
                    0       & h^{ab}\\
                  \end{pmatrix}.
\end{equation}
To recover the generalized metric in the form \pref{dft_gm} the following doubled vielbein can be used
\begin{equation}\label{dft_stdb}
 	\left(E_A{}^M\right) = 
		\begin{pmatrix}
						e_a{}^i	& e_a{}^j B_{ij} \\
						0			& e^a{}_i \\
           \end{pmatrix},
\end{equation}
with $g_{ij} = e^a{}_i e^b{}_j h_{ab}$. There is a local $O(1,D-1)\xX O(1,D-1)$ Lorentz symmetry preserving this gauge choice,
\begin{equation}\label{dft_llt}
    \delta E_A{}^M = {\L_A}^B E_B{}^M,
\end{equation}
with $\L_{AB} = -\L_{BA}$ and $\L_{A}{}^C\cH_{CB} = -\L_{B}{}^C\cH_{AC}$. Due to the choice of constant $\h_{AB}$, the flattened derivative of a vielbein is antisymmetric in its two last indices
\begin{equation}\label{gf_latcon}
	\O_{ABC}  = E_A{}^M\left(\del_M{E_B}^N\right) {E_C}_N = -\O_{ACB}.
\end{equation}
Under a local $O(1,D-1)\xX O(1,D-1)$ transformation this object has a non-covariant variation and can be used to define a covariant derivative. $\O_{ABC}$ is invariant under global $O(D,D)$ coordinates rotations since all indices are properly contracted with $\h^{MN}$. Under the local gauge symmetry \pref{dft_symm} the variation of the vielbein is 
\begin{equation}
	\hat\d_\z E_A{}^M = \hat\cL_\z E_A{}^M 
		= \z^N\del_N E_A{}^M + \del^M \z_N E_A{}^N - \del_N \z^M E_A{}^N.
\end{equation}
This defines the D-bracket of the vector $\z$ with $E_A$, which reduces to the Dorfmann bracket when $\tld\del^m ...= 0$. Objects that will play an important role are the C- and the D-brackets of two vielbeins $E_A{}^M$. Due to the constraint \pref{gf_gbeins} they are equal and, $E_A{}^M$ being invertible, we write
\begin{equation}\label{gf_cdbrack}
 \comm{E_A}{E_B}_{(C)}^M =
    \comm{E_A}{E_B}_{(D)}^M= F_{AB}{}^C E_C{}^M.
\end{equation}
The structure functions are then completely antisymmetric 
\begin{equation}\label{gf_fdef}
 F_{ABC} = \Omega_{ABC} +\Omega_{CAB}+\Omega_{BCA} = 3\O_{[ABC]}.
\end{equation}
$\O_{ABC}$ is not well behaved under the generalized coordinates transformations, however, its antisymmetric part $F_{ABC}$ transforms as a scalar when the strong constraint holds. Another scalar object can be built out of the dilaton and the trace of $\O_{ABC}$
\begin{equation}\label{gf_dder}
 	\tld \O_A = 2 E_A{}^M\del_M d + \O^B{}_{BA}
\end{equation}
Using \pref{gf_gbeins} and \pref{gf_latcon}, the generalized Ricci scalar $\cR$ can be rewritten as
\begin{equation}
\begin{split}
	\cR = &2\cH^{AB} E_A{}^M\del_M\tld\O_B 
		-	\cH^{AB}\tld\O_A\tld\O_B 
		+	\frac{1}{4} \cH^{AB} F_{ACD}F_{B}{}^{CD}\\
		&-	\frac{1}{12}\cH^{AB}\cH^{CD}\cH^{EF}F_{ACE}F_{BDF}
        -\half\cH^{AB}\O^{CD}{}_A\O_{CDB}.
\end{split}
\end{equation}
Using the strong constraint the last term vanishes, after a partial integration the action \pref{dft_raction} reads
\begin{equation}\begin{split}\label{gf_action}
 S 	= \int d^{2D}X\, e^{-2d}\bigg( 
        &\frac{1}{4} \cH^{AB} F_{ACD}F_{B}{}^{CD} 
		-\frac{1}{12}\cH^{AB}\cH^{CD}\cH^{EF}F_{ACE}F_{BDF} 
		+\cH^{AB}\tld\O_A\tld\O_B \bigg).
\end{split}\end{equation}
This rewriting of the DFT action is equivalent to the action of \cite{csw11}, which is a gauged fixed version of this one.

\subsection{Equations of motion}

To derive the equations of motion from the action \pref{gf_action} a Lagrange multiplier $L_{AB}$ has to be introduced to ensure that the vielbein $E_A{}^M$ is an element of $O(D,D)$. If the variation of the action is 
\begin{equation}\label{gf_lmultsvar}
	\d S = \int d^{2D}X e^{-2d} K_{AB} \D^{AB}
\end{equation}
where $\D^{AB} = \d E_A{}^M \,E_{BM}$ then, due to the presence of the Lagrange multiplier $L_{AB}$, the equations of motion involve only the antisymmetric part
\begin{equation}\label{gf_eom}
	K_{[AB]} = 0.
\end{equation}
For the action \pref{gf_action} one gets
\begin{equation}\label{gf_eomk}
	K_{[AB]} = \half(\tld \O_C - E_C{}^M\del_M)Z^C{}_{AB} 
		+\half Z_{[A}{}^{CD}F_{B]CD} -2\cH_{[A}{}^C E_{B]}{}^M\del_M\tld\O_C,
\end{equation}
where 
\begin{equation}\label{gf_zdef}
	Z_{ABC} = 3\cH_{[A}{}^D F_{BC]D} - \cH_A{}^D\cH_B{}^E\cH_C{}^F F_{DEF}.
\end{equation}
The dilaton equation of motion is simply the vanishing of the generalized Ricci scalar $\cR = 0$.

\subsection{DFT and R-R fields}

Double Field Theory has been extended to include the R-R sector of type II strings \cite{hkz11, hkz11b} (see also \cite{fot99,row10}). The key ingredient is the supersymmetric pseudo-action of \cite{bkor01}, where both the R-R fields and their magnetic duals are present, and packaged in a polyform 
\begin{equation}\label{rr_poly}
    F = \sum_p F_p 
    = \sum_p \frac{1}{p!}F_{i_1...i_p} dx^{i_1}\w...\w dx^{i_p},
\end{equation}
where $p$ is either odd (IIB) or even (IIA). The action is the square of the total field strength $F$ and has no topological term, it is supplemented with a self-duality relation imposed to get the right equations of motion
\begin{equation}\label{rr_pact}
    S = \frac{1}{4} \int \ast F\w F ,\qquad F = \ast \s(F),
\end{equation}
where $\s$ is an involution reversing the order of the differentials $dx^i$. The field strength descends from a gauge potential $C$ which contains all the electric and magnetic potentials
\begin{equation}\label{rr_fs}
    F = (d+H\w)C,\qquad (d+H\w)F=0.
\end{equation}
The twisted exterior derivative $d+H\w$ is nilpotent due to the Bianchi identity of the NS-NS three-form $dH=0$. 

Being a sum of even or odd degree differential forms, the total field strength $F$ gives a chiral or antichiral spinorial representation of $O(D,D)$ and transforms as an $O(D,D)$ spinor under T-duality. It is therefore natural to consider the R-R field strength and potential to be $O(D,D)$ spinors in DFT. An action for spinorial R-R fields and its coupling to the generalized metric were constructed in the articles cited above. We do not strictly follow \cite{hkz11} here but rather reformulate these ideas in a way which is more adapted to the vielbein formulation. We define the R-R field strength $\cF$ to be an $O(D,D)$ Majorana spinor transforming under the local $O(1,D-1)\xX O(1,D-1)$ as
\begin{equation}\label{rr_llt}
    \d \cF = \half\L_{AB}\C^{AB}\cF
\end{equation}
with the $O(D,D)$ gamma matrices $\C^A=(\C^a,\C_a)$ and with the convention
\begin{equation}\label{rr_cliff}
    \acom{\C^A}{\C^B} = \h^{AB}.
\end{equation}
Due to the signature of $O(D,D)$, one can choose real gamma matrices $\C^A$ with the property $\C^a =(\C_a)^T$, the Majorana condition becoming a reality condition. The self-duality condition is implemented as
\begin{equation}\label{rr_dual}
    \cF = H \cF,\qquad H =(\C^0 - \C_0)(\C^1 + \C_1) ...(\C^D +\C_D),
\end{equation}
where $H$ squares to the identity for $D = 10$. The spinor $\cF$ can be written in term of the usual R-R field strengths $F_{i_1..i_p}$ as \footnote{The relation of this fields strength to the potential $\ch$ in \cite{hkz11} is $\cF=e^d \bbE \,\slashed \del \ch$ where $\bbE$ is the $Spin(D,D)$ representative of $E_A{}^M$. This change of basis generalizes the $A$- and $C$-basis for the R-R fields.}
\begin{equation}\label{rr_exp}
    \cF = \sum_p \frac{e^\f}{p!} F_{i_1...i_p}\, e_{a_1}{}^{i_1}...e_{a_p}{}^{i_p}\, \C^{a_1 ... a_p}\ket{0},
\end{equation}
where $\ket{0}$ is a Clifford vacuum annihilated by $\C_a$ and where the dilaton pre-factor has been introduced for a later convenience. Using the Clifford algebra \pref{rr_cliff}, the self-duality relation \pref{rr_dual} is equivalent to the one in \pref{rr_pact}. As said before, $\O_{ABC}$ has a non-covariant variation under $O(1,D-1)\xX O(1,D-1)$ local transformations and a covariant derivative commuting with the transformation \pref{rr_llt} can be built
\begin{equation}\label{rr_covd}
    \nabla_A = e^d(E_A{}^M\del_M - \frac{1}{2} \O_{ABC} \C^{BC})e^{-d}.
\end{equation}
The dilaton factor is needed for the covariant derivative of a spinor to transforms as a scalar under doubled diffeomorphisms provided the spinor itself transforms as a scalar. The associated Dirac operator is
\begin{equation}\label{rr_dop}
    \dop = \slashed \del -\slashed F -\half \tld {\slashed \O} 
           = \C^A E_A{}^M\del_M
             -\frac{1}{6}\C^{ABC} F_{ABC} 
             -\half \C^A \tld \O_A
\end{equation}
where $F_{ABC}$ and $\tld \O_A$ are defined as before. With a little algebra it can be shown that this operator is nilpotent when the strong constraint is assumed for all the fields. Using this nilpotency property we can rewrite \pref{rr_fs} as
\begin{equation}\label{rr_fsn}
    \cF = \dop \cA,\qquad \dop \cF = 0,
\end{equation}
where $\cA$ is the potential, the field strength $\cF$ being invariant under gauge transformations $\d \cA = \dop \ch$. To recover the original Bianchi identity $(d+H\w)F=0$ some assumptions have to be made, the doubled vielbein $E_A{}^M$ is expressed as in \pref{dft_stdb} with no dependence in $\tld x_i$ for all fields. With these assumptions $\dop \cF=0$ translates into $(d+H\w)F=0$. An $O(1,D-1)\xX O(1,D-1)$ invariant action can be written
\begin{equation}
	S = \frac{1}{4}\int d^{2D}X e^{-2d} \cF^T B\cF, \qquad B = (\C^0 + \C_0)(\C^0-\C_0),
\end{equation}
where $\cF$ is given by \pref{rr_fsn}. Using \pref{rr_exp}, $\ket{0}^T = \bra{0}$ and $\braket{0}{0}=1$ it reduces to the action \pref{rr_pact}.

As said before, the strong constraint implies the nilpotency of the Dirac operator but, as we shall see in the purely NS-NS compactified case, the converse is not true in general. The square of the operator \pref{rr_dop} acting on the spinor $\cA$ vanishes up to terms involving derivatives contracted between themselves
\begin{equation}\label{rr_nil}
\begin{split}
    \dop^2 \cA &= e^d\del^M\del_M \left( e^{-d} \cA\right) 
                -e^d \C^{BC}\,\O^A{}_{BC} E_A{}^M\del_M\left( e^{-d} \cA \right)\\
                &+\left(-\frac{1}{4}\C^{AB}\,  (\del^M\del_M E_A{}^N)E_{BN}
                +\frac{1}{16}\C^{BCDE} \, \O^A{}_{BC}\O_{ADE} 
                -\frac{1}{16}\O^{ABC}\O_{ABC}\right) \cA=0,
\end{split} 
\end{equation}
and do not by itself implies the strong constraint for all products of fields, e.g. $\O_{A[BC}\O^A{}_{DE]}$ can be vanishing even if does not without the antisymmetrisation. It is interesting that $\dop \cF=0$, with $\cF$ now taken to be a constant internal flux in some compactification, reduces to the non-geometric tadpole cancellation condition of \cite{stw05,stw06}, for the unimodular case $\tld\O_A=0$.

\section{Scherk-Schwarz reduction of DFT}

In this section we compare half-maximal four-dimensional gauged supergravity with a Scherk-Schwarz reduction of DFT. The immediate advantage of the doubled formalism over the standard geometric one is that non-geometric fluxes are built-in right from the beginning as a consequence of the T-duality invariance. The drawback is that finding a general Ansatz for the generalized metric for a given set of fluxes, satisfying all the required constraints, is not known in the general case.

\subsection{Sch\"on and Weidner \texorpdfstring{$\cN=4$}{N=4} four-dimensional gauged supergravity}

The $\cN=4$ supergravity theory with $n$ additional vector multiplets in four dimensions has a rigid duality group which is $SL(2,\bbR)\xX SO(6,n)$. The $6+n$ vectors implement a $U(1)^{6+n}$ local abelian gauge symmetry. The only deformations of this theory preserving the original supersymmetry are gaugings, where a subgroup of the duality group is promoted to a local gauge symmetry, the original abelian vectors fields becoming the non-abelian gauge fields of this local symmetry. The most general gauging of $\cN=4$ four-dimensional supergravity has been constructed in \cite{swe06} using the embedding tensor formalism of \cite{wst05}. The bosonic field content is constituted of a vierbein $e^a{}_\m$, scalars living in the coset 
\begin{equation}
    \frac{SL(2,\bbR)}{SO(2)} \xX \frac{SO(6,n)}{SO(6)\xX SO(n)}
\end{equation}
and a pair of electric $A_{\m}^{+A}$ and magnetic $A_{\m}^{-A}$ gauge fields, so $A_{\m}^{\a A}$ is in the product of fundamentals of $SL(2,\bbR)$ and $SO(6,n)$. The coset $SL(2,\bbR)/SO(2)$ is parameterized by an axion $a$ and a dilaton $\f$, packaged either in a complex scalar $\t = a + ie^{-2\f}$ transforming by M\"obius transformations under $SL(2,\bbR)$ or in a symmetric matrix $M_{\a\b}$ transforming linearly (the reader is referred to the original paper \cite{swe06} for conventions). ${SO(6,n)}/{SO(6)\xX SO(n)}$ is parameterized by a matrix $\cM_{AB}$ transforming as an ${SO(6,n)}$ symmetric tensor. In addition there are auxiliary fields ensuring that only half of the electric/magnetic gauge fields propagate. The gauging of the theory is encoded in the embedding tensor, splitted in representations of $SL(2,\bbR)\xX SO(6,n)$ as $\x_{\a A}$ and $f_{\a ABC}$, completely antisymmetric in $ABC$. For consistency, these gauging parameters are constrained by a set of linear and quadratic relations. Gaugings with $\x_{-A} = f_{-ABC}=0$ are called electric gaugings while those with $\x_{\a A}=0$ are said to be unimodular. The scalar potential reads\footnote{Our conventions differs from those in \cite{swe06} by factors of two in the metric and two-form, namely $g_{\m\n}^{here} = 2 g_{\m\n}^{there}$ and $B_{\m\n}^{here} = \half B_{\m\n}^{++\,there}$}
\begin{equation}\label{dss_n4sp}
\begin{split}
	V = \frac{1}{4}\Bigg(&f_{\a ABC} f_{\b DEF} \cM^{\a\b}
		\bigg[ \frac{1}{3}\cM^{AD}\cM^{BE}\cM^{CF} 
			+\Big(\frac{2}{3}\h^{AD} - \cM^{AD}\Big)
				\h^{BE}\h^{CF} \bigg] \\
	&-\frac{4}{9}f_{\a ABC} f_{\b DEF} \e^{\a\b}\cM^{ABCDEF} 
	 +3\x_\a^M\x_\b^N\cM^{\a\b}\cM_{AB}\Bigg)
\end{split}
\end{equation}
where $\cM^{ABCDEF}$ is a scalar-dependent antisymmetric tensor. Only electric gaugings will be considered here. The $SL(2,\bbR)$ indices are dropped $f_{+ABC} = f_{ABC}$, $\x_{+ A} = \x_{A}$ and the vectors are renamed according to their electric $A_\m^A = A_\m^{+A}$ or magnetic $X_\m^A = A_\m^{-A}$ nature. For these electric gaugings, the dilaton appears in the potential through $\cM^{++} = e^{2\f}$ while the axion has no potential term. In order to match the higher dimensional theory, we set $n=6$. For the particular case of an electric gauging, the constraints for the embedding tensor are summarized by
\begin{equation}\label{dss_etcs}
\begin{split}
    \x^A \x_A           &= 0,\\
    \x^A f_{ABC}        &= 0,\\
    f_{[AB}{}^E f_{C]DE}&= \frac{2}{3}f_{[ABC}\x_{D]},
\end{split}
\end{equation}
in the unimodular case they reduce to a simple Jacobi identity for $f_{ABC}$. The full action for an electric gauging is
\begin{equation}\label{dss_egsg}
\begin{split}
	S = \int \bigg(&R\ast1 
                    -\frac{\ast \cD\t\w \cD\bar\t}{2(\Im\t)^2} 
                    +\frac{1}{8} \ast\cD\cM_{AB} \w\cD\cM^{AB}\\
		            &- \frac{\Im\t}{2} \cM_{AB} \ast F^A\w F^B  
                    + \frac{\Re\t}{2} \h_{AB}\, F^A\w F^B\\
                    &+ \half A^A\w dA_A\w X
                    +\frac{\hat f_{ABE}\hat f_{CD}{}^E}{8} A^A\w A^B\w A^C\w X^D \\
                    &+ \half  \x_A\, B\w\big(dX^A - \hat f_{BC}{}^A A^B \w X^C  \big)
		            - V(\cM)\ast1	\bigg)
\end{split}
\end{equation}
where $A= A^A\x_A$, $X = X^A \x_A$ are shorthands for the connections gauging axion-dilaton transformations and where
\begin{equation}\label{dss_fhat}
   \hat f_{ABC} = f_{ABC} - \x_{[A}\h_{C]B} -\frac{3}{2}\x_B\h_{AC}.
\end{equation}
The covariant derivatives and field strengths entering the action are 
\begin{equation}\label{dss_egsgdef}
\begin{split}
    \cD\t        &= d\t + X + A\t,\\
    \cD\cM_{AB}  &= d\cM_{AB} + 2A^C f_{C(A}{}^D\cM_{B)D} 
                    + A_{(A}\cM_{B)C}\x^C - \x_{(A}\cM_{B)C}A^C, \\
    F^A          &= dA^A - \half f_{BC}{}^A A^B\w A^C 
            -\half A\w A^A + \x^A B.
\end{split}
\end{equation}
For electric gaugings, using the identities \pref{dss_etcs} and \pref{dss_fhat}, one sees that the magnetic gauge field $X^A$ enters the action only through $X$, i.e. contracted with the parameter $\x_A$. The gauge field $X$ corresponds to the gauging of the axion shift $a \rightarrow a +c$ symmetry in $SL(2,\bbR)$ and the gauge field $A$ to the gauging of the scaling symmetry of the dilaton. The covariant derivatives of the axion and the dilaton are 
\begin{equation}
\begin{split}
    \cD a   &= da + X + Aa, \\
    \cD \f  &= d\f -\half A.
\end{split}
\end{equation}
To make contact with the higher dimensional theory the two-form has to be promoted to a propagating field \cite{dpp07}, however, the procedure is different if the gauging is unimodular $\x_A=0$ or not. 

In the unimodular case, the magnetic vector field $X$ and the two-form $B$ disappear of the action. The axion is massless and couples to the electric gauge fields through a Peccei-Quinn term
\begin{equation}\label{dss_adual}
    S = \int \bigg(-\frac{e^{4\f}}{2} \ast da\w da + \frac{a}{2}F^A\w F_A + ...\bigg).
\end{equation}
Provided a Lagrange multiplier three-form $H$ and a new term in the action $-H\w(f-da)$ are introduced, $f = da$ can be treated as independent of $a$. The equation of motion for $a$ yields a Bianchi identity for the three-form
\begin{equation}\label{dss_adualbi}
    dH =-\half F^A\w F_A.
\end{equation}
Imposing this relation in the action and partial integrating the dependency in the scalar $a$ drops. The equation of motion for $f$ is then the duality relation $H = -e^{4\f}\ast f$, so integrating $f$ out one gets a quadratic action for $H$, supplemented with its Bianchi identity
\begin{equation}\label{dss_gsgelecact}
\begin{split}
	S = \int  \bigg(&R\ast1 -2\ast d\f\w d\f - \frac{e^{-4\f}}{2}\ast H\w H\\
		& + \frac{1}{8} \ast\cD\cM_{AB} \w\cD\cM^{AB}- \frac{e^{-2\f}}{2} \cM_{AB} \ast F^A\w F^B - V(\cM)\ast1	\bigg).
\end{split}
\end{equation}

The non-unimodular case is somehow different due to the presence of mass terms for the gauge field $X$ and for the two-form $B$. Since the connection $X$ is non-vanishing, one can use the local shift symmetry of the axion to gauge it away. The action is then 
\begin{equation}\label{dss_egsgmassgf}
	S = \int \bigg(-\frac{e^{4\f}}{2}\ast X\w X - H\w X + ...\bigg),
\end{equation}
where terms not containing the gauge field $X$ have been omitted and where
\begin{equation}\label{dss_hnonuni}
    H = dB -A\w B- \half A^A\w dA_A +\frac{1}{6}f_{ABC}A^A\w A^B\w A^C.
\end{equation}
Integrating $X$, the action for a non-unimodular electric gauging becomes
\begin{equation}\label{dss_gsgnonuniact}
\begin{split}
	S = \int  \bigg(&R\ast1 
                            - 2\ast \cD\f\w \cD\f 
                            - \frac{e^{-4\f}}{2}\ast H\w H\\
                            &+\frac{1}{8} \ast\cD\cM_{AB} \w\cD\cM^{AB}
		                     -\frac{e^{-2\f}}{2} \cM_{AB} \ast F^A\w F^B 
                            - V(\cM)\ast1	\bigg).
\end{split}
\end{equation}
This action is invariant under the gauge transformations\footnote{In the original paper, a parameter $\X = 2(\l - A^A\L_A)$ is used.}
\begin{equation}\begin{split}\label{dss_gxf}
    \d \f       &= \half \L  ,  \\
	\d A^A  	&= d \L^A - f_{BC}{}^{A}A^B\L^C - \x^A \l
                +\half\left( A^A \L - A\L^A +\x^A A^B\L_B\right), \\
	\d B	    &= d\l - \half A\w \l -\half d\L^A \w A_A + \L B\\
	\d \cM_{AB} &= -2 \L^C f_{C(A}{}^D\cM_{B)D} ,
                + \x_{(A}\cM_{B)C} \cL^C  - \cL_{(A}\cM_{B)C}\x^C,
\end{split}\end{equation}
where $\L = \L^A\x_A$.

\subsection{Scherk-Schwarz reduction of DFT}

The doubled coordinates are split as $4$ external coordinates $x^\m$, $4$ dual coordinates $\tld x_\m$ and $6+6$ internal doubled ones $Y^M =(y^m,\tld y_m)$:
\begin{equation}\label{dss_splitt}
	\hat X^{\hat M} = \left( x^\m, \tld x_\m, Y^M \right),
\end{equation}
where the hats denote $(10+10)$-dimensional quantities and indices. The flat space indices $\hat A$ are split as $V^{\hat A} = \left(V^a,V_a,V^A\right)$. All fields and gauge parameters are chosen to be independent of $\tld x_\m$ such that the external part of the strong constraint is trivial. Residual symmetries of the compactified theory should be identified with the symmetries of the effective theory. Generalized diffeomorphisms of the $x^\m$ coordinates generate external diffeomorphisms in the effective theory, $\tld x_\m$ shifts generate the gauge transformations of the two-form and $Y^M$ transformations are identified with gauge transformations of the vectors. The simplest procedure is to first consider a Kaluza-Klein (KK) reduction of the theory where the fields and gauge parameters do not depend on the internal coordinates. Then, by looking at the variation of the components of $\hat E_{\hat A}{}^{\hat M}$ under a generalized diffeomorphism with parameter $\hat\z^{\hat M}(x)$ one can identify the effective fields ($e_a{}^\m$,$A_{\m}^A$,...) in $\hat E_{\hat A}{}^{\hat M}$. The Scherk-Schwarz reduction is then done by keeping the modes of the KK reduction but with twists depending on the internal coordinates.  For the Scherk-Schwarz reduction, the gauge parameters of the residual doubled diffeomorphisms are factorized the following way
\begin{equation}\label{dss_rdd}
	\hat\z^{\hat M} = \left( \x^\m(x),\, e^\c \l_\m(x),\, e^\frac{\c}{2}\L^A(x)E_A{}^M(Y) \right),
\end{equation}
where $\c$ is a warp factor depending only on the internal coordinates and where $E_A{}^M$ is a doubled internal vielbein. An Ansatz for the vielbein $\hat E_{\hat A}{}^{\hat M}$ should be expressed in function of the effective fields, such that its variation under generalized diffeomorphisms generated by \pref{dss_rdd} matches the variation of the latter under the gauge transformation \pref{dss_gxf}. By inspection of the transformation of $\hat E_{\hat A}{}^{\hat M}$ under \pref{dss_rdd}, the following Ansatz
\begin{equation}\begin{split}\label{dss_ansatz}
	\hat E_a{}^\m 		&= e^{-\f -\frac{\c}{2}}    e_a{}^\m(x) \\
	\hat E^{a\m}		&= 0\\
	\hat E_A{}^\m		&= 0\\
	\hat E_{a\m} 		&= e^{-\f+\frac{\c}{2}}e_a{}^\n
                         \left(B_{\m\n}(x)  -\half A_\m{}^A A_{\n A}  \right) \\
	\hat E^{a}{}_\m		&= e^{\f+\frac{\c}{2}}   e^{a}{}_\m(x) \\
	\hat E_{A\m}		&= e^{\frac{\c}{2}}\F_A{}^B(x)  A_{\m B}(x) \\
	\hat E_a{}^M 		&= - e^{-\f}e_a{}^\m(x)  A_{\m}{}^A  E_{A}{}^M(Y)\\
	\hat E^{a M}		&= 0\\
	\hat E_A{}^M		&= \F_A{}^B(x) E_{B}{}^M(Y)\\
\end{split}\end{equation}
is found to have the correct gauge transformations. Its dependence in the internal dimensions $Y^M$ is completely factorized in the internal vielbein $E_A{}^M$ and the warp factor $\c$. The full doubled dilaton is chosen to be 
\begin{equation}\label{dss_dil}
	\hat d = -\frac{1}{4} \log{\det{g}} - \f(x) + d(Y)
\end{equation}
where $g = -\det g_{\m\n}$ and $g_{\m\n} = e^a{}_\m e^b{}_\n \,h_{ab}$. Provided $\F_A{}^B$ and $E_A{}^M$ are elements of $O(6,6)$ the Ansatz satisfies the constraint
\begin{equation}\label{dss_ecsrt}
	\hat E_{\hat A}{}^{\hat M} \hat E_{\hat B\hat M} 
        = \hat \h_{\hat A\hat B}.
\end{equation}
Calculation of the components of $\hat F_{\hat A\hat B\hat C}$ and $\hat {\tld\O}_{\hat A}$ is then performed. The components of $\hat{\tld \O}$ are 
\begin{equation}\label{dss_otldcomp}
\begin{split}
	\hat{\tld \O}_a	&=  e^{-\f-\frac{c}{2}}\bigg(\t_{ab}{}^b - e_a{}^\m
        \left( \del_\m \f + e^\frac{\c}{2} A_\m{}^A\tld\O_A\right)\bigg), \\
	\hat{\tld \O}^a 	&= 0, \\
	\hat{\tld \O}_A 	&= \F_A{}^B\tld\O_B ,
\end{split}
\end{equation}
where
\begin{equation}
 	\tld\O_A		= 2E_A{}^M\del_M d + \O^B{}_{BA}.
\end{equation}
Components of $\hat F$ are then computed, for three external lower indices we find
\begin{equation}\label{dss_3form}
\begin{split}
	\hat F_{abc} = - e^{-3\f-\frac{\c}{2}} e_a{}^\m e_b{}^\n e_c{}^\r
		\bigg\{ &3 \del_{[\m}B_{\n\r]} -3A_{[\m}{}^A B_{\n\r]}\, e^\frac{\c}{2} E_A \c\\
				&-3 \del_{[\m}A_{\n}{}^A A_{\r] A} 
				+A_\m{}^A A_\n{}^B A_\r{}^C e^\frac{\c}{2} F_{ABC} \bigg\},
\end{split}
\end{equation}
where $F_{ABC}$ is defined as in the previous section but for the internal vielbein
\begin{equation}
	F_{ABC} = 3 (E_{[A}{}^M\del_ME_B{}^N)E_{C]}.
\end{equation}
This component thus contains the full three-form field strength $H$ with Chern-Simons term given in \pref{dss_hnonuni}
\begin{equation}\label{dss_hcs}
	\hat F_{abc} = - e^{-3\f-\frac{\c}{2}} e_a{}^\m e_b{}^\n e_c{}^\r H_{\m\n\r},
\end{equation}
provided the following identifications are made
\begin{equation}\label{dss_paramid}
\begin{split}
	e^\frac{\c}{2}E_A\c		 &=	\x_A,\\
	e^\frac{\c}{2}F_{ABC}	 &= f_{ABC}.
\end{split}
\end{equation}
This field strength enters the action \pref{gf_action} through the $\hat\cH\hat\cH\hat\cH\hat F\hat F$ term 
\begin{equation}\label{dss_reduxH}
	-\frac{1}{12}\int d^{2D}X \sqrt{-g} e^{-4\f - \c - 2d}  
		g^{\m\n}g^{\r\s}g^{\l\t} H_{\m\r\l}H_{\n\s\t},
\end{equation}
reproducing the correct kinetic term in \pref{dss_gsgnonuniact}. The component with one internal index and two external indices is proportional to the non-abelian Yang-Mills field strength for the gauge fields
\begin{equation}
\begin{split}
	 	\hat F_{abC} &= -e^{-2\f -\frac{\c}{2}} e_a{}^\m e_b{}^\n \F_C{}^D F_{\m\n D} \\
					 &= -e^{-2\f -\frac{\c}{2}} e_a{}^\m e_b{}^\n \F_C{}^D 
								\bigg\{ 2\del_{[\m}A_{\n]D} 
										- f_{DAB} A_\m{}^A A_\n{}^B 
										- A_{[\m}A^A_{\n]C} 
										+ \x_C B_{\m\n} \bigg\}.
\end{split}
\end{equation}
Once again, the $\hat\cH\hat\cH\hat\cH\hat F\hat F$ term reduces to the correct kinetic term, provided the symmetric scalar moduli matrix is expressed as
\begin{equation}\label{dss_phim}
	\cM^{AB} = \cH^{CD}\F_C{}^A\F_D{}^B.
\end{equation}
The scalars covariant derivatives are contained in $\hat F_{aBC}$
\begin{equation}
\begin{split}
    \hat F_{aBC} &= e^{-\f-\frac{\c}{2}} e_a{}^\m
        \left(\cD_\m \F_{BA} \right) \F_C{}^A,\\
    \cD_\m \F_{BA} &= \del_\m \F_{BA}
                    +A_\m{}^C f_{CA}{}^D \,\F_{BD}
                    +\half A_{\m A}\, \F_{BD}\,\x^D
                    -\half \x_A\, \F_{BD}\,A_\m{}^D. 
\end{split}
\end{equation}
Using the formula \pref{dss_phim}, the correct kinetic term is recovered. The scalar potential comes from the $\hat F_{ABC}$, $\hat F_{A b}{}^c$ and the $\hat {\tld \O}_A$ components. Plugging those in the action we get
\begin{equation}\label{dss_reduxscal}
\begin{split}
	 \frac{1}{4}\int d^{2D}X e^{2\f -\c - 2 d} \bigg\{
		 \cM^{AB}f_{ACD}f_{B}{}^{CD} 
		 -\frac{1}{3}\cM^{AB}\cM^{CD}\cM^{EF}f_{ACE}f_{BDF}
		 -3 \cM^{AB}\x_A\x_B \bigg\}.
\end{split}
\end{equation}
where we made the identification 
\begin{equation}\label{dss_tldoxi}
	\tld \O_A = -\frac{e^{-\frac{\c}{2}}}{2} \x_A. 
\end{equation}
The gravitational sector is completely contained in $\hat {\tld \O}_a$ and $\hat F_{ab}{}^c$,
\begin{equation}\label{dss_gravfs}
\begin{split}
	\hat{\tld \O}_a	&=  e^{-\f-\frac{c}{2}}\left(\t_{ab}{}^b - \cD_a\f\right), \\
	\hat F_{ab}{}^c &= e^{-\f-\frac{\c}{2}}\left( \t_{ab}{}^c +2\d_{[a}{}^c \cD_{b]} \f\right),\\
	\cD_a\f 		&= e_{a}{}^\m\left( \del_\m\f - \half A_\m \right),\\
	\t_{ab}{}^c 	&= \left(e_{a}{}^\m\del_\m e_{b}{}^\n - e_{b}{}^\m\del_\m e_{a}{}^\n\right)e^c{}_\n.
\end{split}
\end{equation}
After a partial integration one gets the Ricci scalar and the kinetic term for the charged dilaton.

\subsection{Analysis of the constraints}

For the dependence in the internal dimensions to factorize, the gauging parameters, built out of derivatives of the internal doubled vielbein $E_A{}^M(Y)$, the internal dilaton $d$ and the warp factor, must be constant
\begin{equation}\begin{split}\label{dss_ecb}
	f_{ABC}		&= e^\frac{\c}{2}F_{ABC} = cst.,\\
	F_{ABC} 	&= \O_{ABC} + \O_{CAB} + \O_{BCA} = F_{[ABC]},\\
	\x_A		&=  e^\frac{\c}{2} E_A{}^M\del_M \c 
        = -2\, e^\frac{\c}{2}(2E_A{}^M\del_M d +\O^B{}_{BA} )= cst.	
\end{split}\end{equation}
This places severe constraints on the internal fields that should be analyzed, we will assume them for the moment and treat this problem for examples in the last section. With these assumptions, the dependence in the internal coordinates in the reduced action factorizes as
\begin{equation}
	\int d^{12}Y\, e^{-2d -\c}.
\end{equation}
It is easy to show, using the last line of \pref{dss_ecb} and the transformation of the dilaton, that this object is invariant under gauge transformations parameterized by $\hat \z^M$. A question that one may ask is, assuming these constraints for constancy of the parameters, are there additional constraints needed to ensure the closure conditions \pref{dss_etcs}. The answer is that, provided the strong constraint for $E_A{}^M$, $\c$, and $d$ is assumed, \pref{dss_etcs} holds. We first show that
\begin{equation}
	E_{[A}{}^M\del_MF_{BCD]} 
        = 3\O_{[AB}{}^E\O_{CD]E} + 3\O^E{}_{[AB}\O_{CD]E}.
\end{equation}
With this identity, and using the properties of $f_{ABC}$, it is easy to demonstrate that the modified Jacobi identity 
\begin{equation}
	f_{[AB}{}^E f_{C]DE} = \frac{2}{3}f_{[ABC}\x_{D]}
\end{equation}
holds provided an additional constraint is imposed
\begin{equation}\label{dss_wrc}
	\O_{E[AB} \O^E{}_{C]D} = 0.
\end{equation}
Notice that the strong constraint for the internal frame $E_A{}^M$ reads
\begin{equation}\label{dss_sce}
	\O^E{}_{AB} \O_{ECD} = 0.
\end{equation}
It seems therefore that the modified Jacobi, needed for the effective action to have the full non-abelian symmetry, is weaker than the strong constraint. By looking carefully at the expressions for $E^{AM}\del_M f_{ABC}$ and $E_{[B}{}^M\del_M \x_{C]}$, for $f_{ABC}$ and $\x_A$ given by \pref{dss_ecb}, we find that the closure relation
\begin{equation}
	\x^A f_{ABC} = 0
\end{equation}
holds only provided the strong constraint between the dilaton, the warp factor and vielbein holds
\begin{equation}
	\O_{ABC}\,  E^{AM}\del_M d= \O_{ABC}\,  E^{AM}\del_M \c= 0
\end{equation}
and provided a constraint, which is actually weaker than the weak level-matching constraint $\del^M\del_M\, E_{A}^N =0$, holds
\begin{equation}
	(\del^M\del_M\, E_{[A}^N) E_{B]N} = 0.
\end{equation}
Finally, for the $\x^A\x_A = 0$ relation to hold, one should impose
\begin{equation}
	\del^M\c\, \del_M \c = \tld\O^A\, \tld\O_A = 0.
\end{equation}
The strong and the weak level-matching constraints for the internal vielbein are sufficient but not necessary conditions for the closure relations to hold. These constraints are needed in the uncompactified DFT for the generalized diffeomorphisms to be a symmetry of the action. When compactified on some background, these diffeomorphisms are broken up to the subgroup of (generalized) isometries of the background. For these residual symmetries to be symmetries of the compactified action some constraints still have to be imposed, but are weaker than the original ones.

\subsection{The \texorpdfstring{$\cN=4$}{N=4} term}

Comparison of the reduced theory with the gauged supergravity shows a discrepancy in the potential, the supergravity having an additional term in the scalar potential
\begin{equation}\label{dss_addterm}
	\tld V = \frac{e^{2\f}}{6} f_{ABC}f^{ABC}.
\end{equation}
The same discrepancy has already been discovered in \cite{amnr11, dgr11} for the embedding of $\cN=4$ in $\cN=8$, yielding two additional quadratic conditions for the embedding tensor, one of those being the vanishing of \pref{dss_addterm} and the other related to the total O-plane/D-brane charge. If it is assumed that a modified Double Field Theory exists such that the strong constraint is not required for the consistency of the theory, other terms with all the symmetries can be added to the original action \pref{gf_action}.
For example 
\begin{equation}\label{dss_news}
	\tld S = -\frac{1}{6} \int d^{2D}X e^{-2\hat d} \hat F_{\hat A\hat B\hat C} \hat F^{\hat A\hat B\hat C}
\end{equation}
precisely gives the required term for both potentials to match but nothing more (a similar term has been found for the heterotic extension of DFT \cite{hkw11b}). Upon Scherk-Schwarz reduction, it leads to an additional contribution to the dilaton potential and does not change the $\hat E_{\hat A}{}^{\hat M}$ equation of motion. Expanding $\hat F_{\hat A\hat B\hat C}$ in \pref{dss_news} and partial integrating it becomes 
\begin{equation}
	\tld S =  \int d^{2D}X e^{-2\hat d}\left\{ -2 \del_{\hat M}\del^{\hat M} \hat d 
							+\hat {\tld\O}^{\hat A}\hat {\tld\O}_{\hat A}
							-\half \hat\O^{\hat A\hat B\hat C}\hat\O_{\hat A\hat B\hat C}  \ \right\},
\end{equation}
and hence, in our Scherk-Schwarz reduction, vanishes when the level-matching constraint for $d$ and the strong constraint for the vielbein are assumed. Looking at the expression for the nilpotency of the Dirac operator\pref{rr_nil} we see that the last term multiplicating the R-R gauge potential is proportional to the additional term in the potential. Hence, if R-R fields are present and the strong constraint holds for all fields except the vielbein, this term has to vanish and we recover the additional condition of \cite{dgr11} on the $\cN=4$ embedding tensor for the theory to be a truncation of $\cN=8$.

\subsection{Equations of motion and vacua}

In this section and for the rest of the paper, we restrict ourselves to the unimodular case $\x_A=0$ and $f_{ABC} = F_{ABC}$. A vacuum solution has to satisfy the equations of motion. With the assumptions of constant $F_{ABC}$  and vanishing $\tld\O_A$, \pref{gf_eom} reduces to
\begin{equation}\label{nsred_bkgeomnc}
	\cH_{[A}{}^G F_{B]CD} F_{EFG}\left(\h^{CE}\h^{DF} - \cH^{CE}\cH^{DF}\right) = 0.
\end{equation}
This can be recast to a more instructive form introducing projectors 
\begin{equation}\label{nsred_biproj}
	P_{\pm\,ABCD} 		= \half(\h_{A(C}\h_{D)B} \pm \cH_{A(C}\cH_{D)B})
\end{equation}
such that \pref{nsred_bkgeomnc} reads
\begin{equation}\label{nsred_bkgeomproj2}
	W^{AB} = P_-^{ABCD}Z_{CD} = 0
\end{equation}
i.e. the Weyl anomaly of the sigma-model of \cite{adp09}, with
\begin{equation}\label{nsred_zdef}
	Z_{AB} = \frac{1}{4} F_{ACD}F_{BEF} P_-^{CEDF}.
\end{equation}
The dilaton equation of motion reads
\begin{equation}\label{nsred_bkgdil}
	-\frac{1}{12}Z^{ABC}F_{ABC}= R^{ext}
\end{equation}
with $Z^{ABC}$ defined in \pref{gf_zdef} and $R^{ext}$ the Ricci scalar of the external space. The potential reads
\begin{equation}\label{nsred_bkgpot}
	V(\cM) = -\frac{1}{12} Z^{ABC}(\cM) F_{ABC}
\end{equation}
where $Z_{ABC}(\cM)$ is defined like $Z_{ABC}$ but with $\cH_{AB}$ replaced by $\cM_{AB}$. The projector 
\begin{equation}\label{nsred_biprojm}
	P_{\pm\,ABCD}(\cM) 	= \half(\h_{A(C}\h_{D)B} \pm \cM_{A(C}\cM_{D)B})
\end{equation}
and $Z_{AB}(\cM)$ are defined by the same replacement. The dilaton equation of motion tells that when $\cM = \cH$ the potential equals the external curvature. To study extrema, $\cM_{AB}$ being a constrained field one has to introduce a Lagrange multiplier 
\begin{equation}\label{dss_lagmult}
	\hat V = V + L^{AB}\left(\cM_{AB} - (\cM^{-1})^{CD}\h_{CA}\h_{DB}  \right).
\end{equation}
Variation with respect to $\cM$ leads to the equation
\begin{equation}\label{nsred_potmin}
	\frac{\del \hat V}{\del \cM^{AB}} = -Z_{AB}(\cM) + 2P_{+\,ABCD}(\cM) L^{CD} =0
\end{equation}
with 
\begin{equation}\label{dss_varpot}
\begin{split}
	-Z_{AB}(\cM) = \frac{\del V}{\del \cM^{AB}} &= - \frac{1}{4}F_{ACD}F_{BEF} 
		\left(\h^{CE}\h^{DF} - \cM^{CE}\cM^{DF}\right)\\
	&= F_{ACD}F_{BEF} P_-^{CEDF}(\cM) =0.
\end{split}
\end{equation}
This says that at an extremum of $V$ the following equation is satisfied
\begin{equation}\label{nsred_vextr}
	P_-^{ABCD}(\cM)Z_{CD}(\cM) =  0
\end{equation}
and therefore, assuming \pref{nsred_bkgeomproj2} and \pref{nsred_bkgdil}, the point $\cM_{AB} = \cH_{AB}$ is an extremum of the potential with external Ricci scalar curvature given by \pref{nsred_bkgdil}.

\section{Doubled Backgrounds}

The assumption that $F_{ABC}$ defined in \pref{dss_ecb} is constant places constraints on the frame $E_A{}^M$ and it is not clear a-priori that a general solution can be found. In the usual, non-doubled, Scherk-Schwarz reduction a large class of solutions to this problem is given by group manifolds and quotients of them by a discrete subgroup. On such geometries the twist matrix factorizing the dependence of the fields in the internal coordinates can be identified with the Maurer-Cartan frame $e_a{}^m$,
\begin{equation}\label{db_grpman}
	e^a T_a = g^{-1}d g, \quad de^a = -\half \t_{bc}{}^a e^b\w e^c, \quad \comm{T_a}{T_b} = \t_{ab}{}^c T_c.
\end{equation}
The functions $\t_{ab}{}^c$ appearing in the reduced theory are therefore identified with the structure constants of the Lie algebra in the particular basis \pref{db_grpman}. Additional assumptions on the potential reduces the choice of group to flat groups and compactness of the internal space requires the identification of points by a discrete subgroup, this class of manifold has often been dubbed `twisted tori' in the string literature.

In the present, doubled, case the situation is somewhat different since the frame $E_A{}^M$ has to belong to $O(6,6)$ and has to satisfy the strong constraint, or at least a weaker constraint ensuring the Jacobi identity of $F_{ABC}$. Furthermore the structure constants are no longer derived from the Maurer-Cartan equation but rather from the C-bracket, the latter reducing to the Maurer-Cartan equation in some cases. Another fundamental feature of the charges $F_{ABC}$ defined this way is that, being totally antisymmetric the Lie algebra they define leaves the metric $\h_{AB}$ invariant and the gauged group in the effective theory is automatically a subgroup of $O(6,6)$.

\subsection{Geometric and three-form fluxes}

Standard Scherk-Schwarz compactifications with three-form flux $H_{abc}$ can be described quite simply within this formalism. The Jacobi identity for the doubled flux $F_{ABC}$ has two non-trivial parts, the Jacobi for the geometric flux $\t_{ab}{}^c$ and a mixed one \footnote{To simplify the notations and to comply with the existing literature we use latin letters $a,b,c...= 1,...,n$ for flat internal non-doubled indices, whereas we used those for flat external indices in the previous section.}
\begin{equation}\label{kms_jaco}
\begin{split}
	\t_{[ab}{}^d\t_{c]d}{}^e 	&= 0 \\
	H_{e[ab}\t_{cd]}{}^e 		&= 0.
\end{split}
\end{equation}
An Ansatz for the doubled vielbein, that is manifestly an $O(n,n)$ matrix and trivially satisfying the strong constraint, can be introduced (see also \cite{allp11})
\begin{equation}\label{kms_ansatz}
 E_A{}^N = {\begin{pmatrix}
						e_a{}^n(y)		& e_a{}^m(y)B_{mn}(y) \\
						0							& e^a{}_n(y) \\
           \end{pmatrix}_A}^N
\end{equation}
with antisymmetric $B_{mn}$ and $m=1,...,n$. Computing the C-brackets of this frame one quickly obtains 
\begin{align}\label{kms_calg}
	\comm{E_a}{E_b} &= H_{abc} E^c + \t_{ab}{}^c E_c \\
	\comm{E^a}{E_b} &= \t_{bc}{}^a E^c \\
	\comm{E^a}{E^b}	&= 0
\end{align}
where $\t_{ab}{}^c$ is given by the Maurer-Cartan equation for $e^a$ and 
\begin{equation}\label{kms_flux}
 H_{abc} = 3 e_{[a}{}^m e_b{}^n e_{c]}{}^p \del_m B_{np}.
\end{equation}
Note that in this case the three terms structure of the C-Bracket is required for the gauge invariance of $H_{abc}$. One still has to find the internal generalized dilaton, by setting
\begin{equation}\label{sss_dil}
	e^{-2d} = \det(e^a{}_m)
\end{equation}
the condition $\tld\O_A = 0$ translates in the unimodularity condition $\t_{ab}{}^b = 0$.

\subsection{Solution with \texorpdfstring{$Q$}{Q}- and \texorpdfstring{$R$}{R}-flux}

Keeping for now only dependence in the coordinates $y^m$, one can find an Ansatz with non-geometric flux $Q_a{}^{bc}$
\begin{equation}\label{dbg_qansatz}
 E_A{}^N = {\begin{pmatrix}
						\d_a{}^n			& 0\\
						\b^{ab}(y)\d_b{}^m	& \d^a{}_n \\
           \end{pmatrix}_A}^N.
\end{equation}
Writting $y^a = \d^a_m y^m$ and $\del_a = \d_a^m \del_m$, only two pieces of $\O_{ABC}$ are non-vanishing
\begin{equation}\label{dbg_qfluxsc}
\begin{split}
		\O_{a}{}^{bc} &= \del_a \b^{bc} \\
		\O^{abc}		&= \b^{ad}\del_{d}\b^{bc}.
\end{split}
\end{equation}
For $Q_a{}^{bc} = F_a{}^{bc}$ to be constant $\b$ should have at most a linear dependence in $y^a$
\begin{equation}\label{dbg_qfluxtwist}
	\b^{ab}(y) = y^c \b_c{}^{ab}.
\end{equation}
The non-trivial Jacobi identities are 
\begin{equation}\label{dbg_qfluxjaco}
\begin{split}
	Q_d{}^{[ab} Q_e{}^{c]d}	&= 0 \\
	R^{e[ab}Q_e{}^{cd]}		&= 0.
\end{split}
\end{equation}
Therefore one find that only $Q$-flux is non-vanishing
\begin{equation}\label{dbg_qflux}
\begin{split}
	Q_a{}^{bc}	&= \b_a{}^{bc}\\
	R^{abc}		&= 3\b^{[a\vert d}\b_{d}{}^{\vert bc]} = 0.
\end{split}
\end{equation}
Introducing a dependence of $\b$ in the dual coordinates $\tld y_m$ is also possible and leads to $R$-flux
\begin{equation}\label{dbg_qrfluxtwist}
	\b^{ab} = y^c \b_c{}^{ab} + \tld y_c \b^{cab}.
\end{equation}
One has to take care in this case of the strong constraint, which is quite strong due to the absence of antisymmetrization. If we impose its weaker version \pref{dss_wrc} instead, only the Jacobi identity remains
\begin{equation}\label{dbg_qrfluxsc}
	\b^{a[bc}\b_a{}^{de]} = 0
\end{equation}
with $Q_a{}^{bc}= \b_a{}^{bc}$ and $R^{abc}= \b^{abc}$.

\subsection{More general twists}

More general twisted background can be constructed \cite{hul04,hre09,dpst07}. As an example consider compactifying one coordinate $y$ and its dual $\tld y$ on circles, the $2(n-1)$ remaining coordinates being labeled with a bar on their index $Y^A = (y,\tld y,Z^{\bar A})$. An antisymmetric matrix $\cN_{AB}$ is chosen with zeros in $y$ and $\tld y$ rows and columns, with $\cN_{\bar A \bar B} = -\cN_{\bar B \bar A}$ for it to be in the Lie algebra of $O(n-1,n-1)$. The Ansatz 
\begin{equation}
 	E_A{}^M = \left(e^{y\, \cN} \right)_A{}^B\, \d_B^M 
\end{equation}
then describes a $2(n-1)$-torus with an $O(n-1,n-1)$ twist over a $\bbT^2$ base. The only non-zero piece of $\O_{ABC}$ is
\begin{equation}
	\O_{y \bar A \bar B} = \cN_{\bar A \bar B}.
\end{equation}
This corresponds to the gauging of a Lie algebra, with generators $T$, $\tld T$ and $T_{\bar A}$
\begin{equation}
\begin{split}
	\comm{T_{\bar A}}{T_{\bar B}} 	&= \cN_{\bar A \bar B} T\\
	\comm{\tld T}{T_{\bar A}}		&= \cN_{\bar A \bar B} T^{\bar B}.
\end{split}
\end{equation}
This example can be generalized to include cases where the strong constraint does not hold but still seems to yield valid gaugings in the effective theory. 
First split internal coordinates $Y^A = \d^A_M Y^M$ into $(Y^I, Z^{\bar A})$ with $I= 1,...,2b$ and $\bar A = 1,...2f$. Then introduce the Ansatz
\begin{equation}
 	E_A{}^M = \left(e^{Y^I\, T_I} \right)_A{}^B\, \d_B^M,
\end{equation}
where the only non-vanishing component in $T_{IBC}$ is $T_{I\bar A\bar B}$. We also require that the generators $T_I$ commute between themselves and generate an $O(f,f)$ subalgebra, $T_{I\bar A\bar B} = - T_{I \bar B \bar A}$. The only non-zero component of $\O_{ABC}$ is in this case
\begin{equation}
	\O_{I\bar A\bar B}=T_{I\bar A\bar B}.
\end{equation}
This is equivalent to the gauging of the algebra
\begin{equation}
\begin{split}
	\comm{T_I}{T_J}			&=0\\
	\comm{T_I}{T_{\bar A}}	&=T_{I\bar A}{}^{\bar B}\, T_{\bar B}\\
	\comm{T_{\bar A}}{T_{\bar B}}	&= T_{I\bar A\bar B}\, T^I.
\end{split}
\end{equation}
and the only non-trivial Jacobi identity is
\begin{equation}\label{dbg_nscjaco}
	T^I{}_{[\bar A\bar B} T_{I\bar C]\bar D} = 0.
\end{equation}
Since the generators $T_I$ commute, a possible choice is to take them to be diagonal 
\begin{equation}
	T_{Ia}{}^b = - T_I{}^b{}_a = \a_{Ia}\d_a^b
\end{equation}
and \pref{dbg_nscjaco} translates into
\begin{equation}
	\a^I{}_a \a_{Ib} = 0,\quad a \neq b.
\end{equation}
One can try to solve the equations of motion for this solution, the vanishing of $W_{IJ}$ implies
\begin{equation}\label{dbg_nscos}
	\a_{Ia} = \pm \cH_{IJ}\a^J{}_a 
\end{equation}
and all others equations are trivial. It is interesting that for these solutions the term \pref{dss_addterm} in the potential is non-zero. Evaluating the potential at $\cM_{AB} = \cH_{AB}$ with the on-shell condition \pref{dbg_nscos} gives
\begin{equation}
	V(\cH) = - e^{2\f} (1 \mp 1) \a^I{}_a\a_{Ib}\d^{ab},
\end{equation}
showing that the additional term in the action is required for this class of solutions to have Minkowski vacua.

\section*{Conclusion}

In this paper the Double Field Theory of \cite{hhz10b} describing the dynamics of the generalized metric of string theory was reformulated in a vielbein formalism. A generalization of the Scherk-Schwarz procedure was developed and accordance with half-maximal gauged supergravity was demonstrated provided a new term is added to the original action. Several classes of examples of doubled backgrounds were presented. It was shown that only a weaker version of the strong constraint on background fields is needed to ensure consistency conditions of the effective theory.

\section*{Acknowledgements}

This work began in collaboration with Nicolas Ambrosetti, Daniel Arnold and Nikolaos Prezas, the author warmly thanks them for all the work done together and for passionate discussions. The author also thanks Matthias Blau and Jelle Hartong for enlightening conversations. Finally the author thanks Jean-Pierre Derendinger for numerous discusions about this work. For some of these tedious computation, the tensor CAS described in \cite{pee06} was found to be very useful. D.G. is supported by the Swiss National Science Foundation.

\bibliographystyle{JHEP}
\bibliography{dftn4.v2}

\end{document}